\documentclass[amsfonts, amssymb, pre, superscriptaddress, notitlepage, twocolumn,11pt,tightenlines]{revtex4-2}

\usepackage{amsmath,url}
\usepackage{graphicx}
\usepackage{CrimsonPro}
\usepackage[letterpaper, margin=20mm]{geometry}
\newcommand{\etal}{\textit{et al.}}

\begin{document}

\title{Networks of climate change: Connecting causes and consequences}

\author{Petter Holme}
\affiliation{Department of Computer Science, Aalto University, Espoo, Finland}
\affiliation{Center for Computational Social Science, Kobe University, Kobe, Japan}

\author{Juan C. Rocha}
\affiliation{Stockholm Resilience Centre, Stockholm University, Sweden}

\begin{abstract}
Understanding the causes and consequences of, and devising countermeasures to, global warming is a profoundly complex problem. Network representations are sometimes the only way forward, and sometimes able to reduce the complexity of the original problem. Networks are both necessary and natural elements of climate science. Furthermore, networks form a mathematical foundation for a multitude of computational and analytical techniques. We are only beginning to see the benefits of this connection between the sciences of climate change and network science. In this review, we cover the wide spectrum of network applications in the climate-change literature---what they represent, how they are analyzed, and what insights they bring. We also discuss network data, tools, and problems yet to be explored.
\end{abstract}

\maketitle

\begin{figure}
  \includegraphics[width=\linewidth]{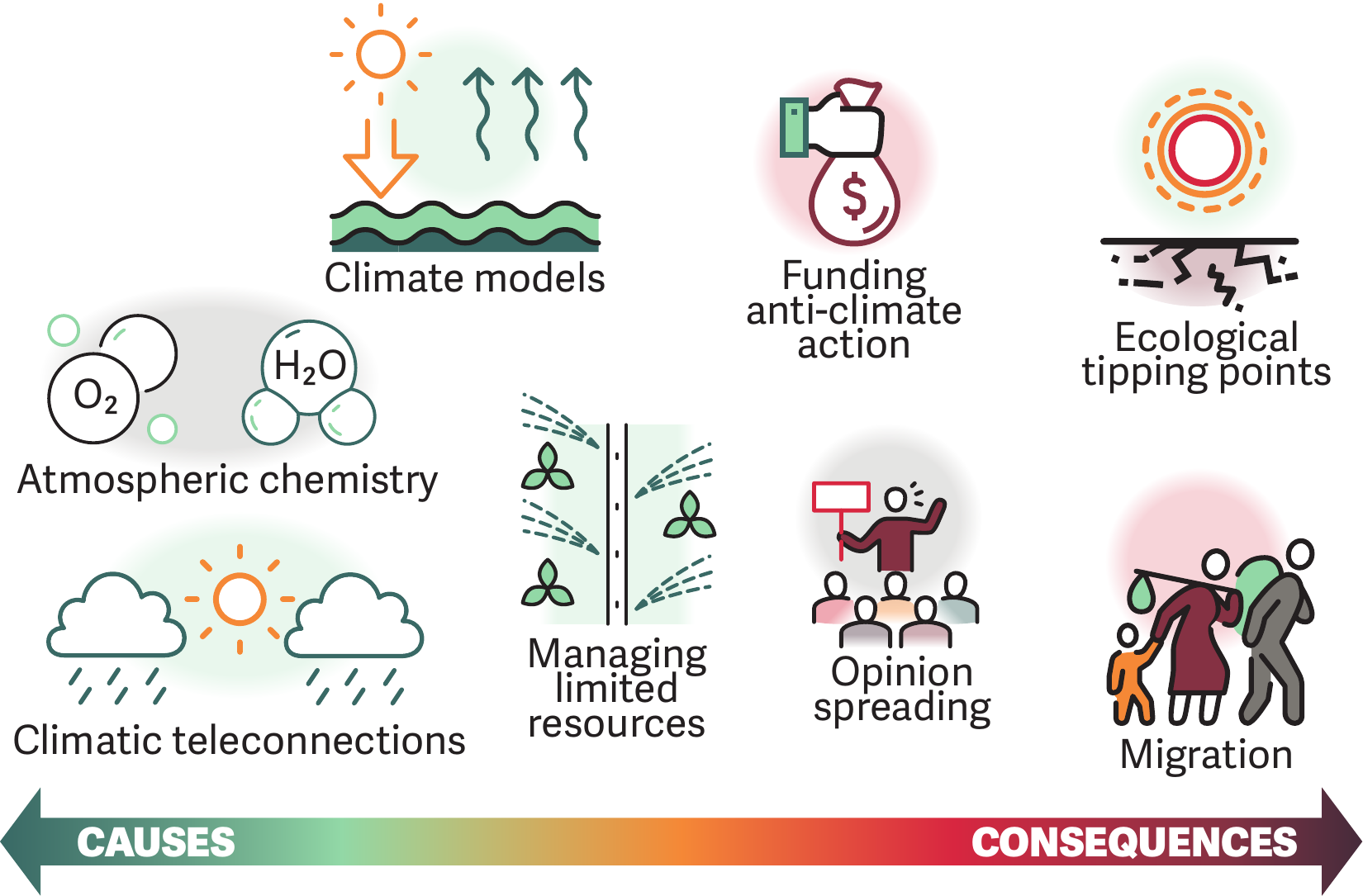}
  \caption{\textbf{Overview of topics related to climate change that have been studied by networks.} Most topics in the science of climate change are causally coupled. Still, in the discourse, some have more the flavor of causes, some are more consequences.}\label{fig:overview}
\end{figure}

\begin{figure*}
  \includegraphics[width=0.8\linewidth]{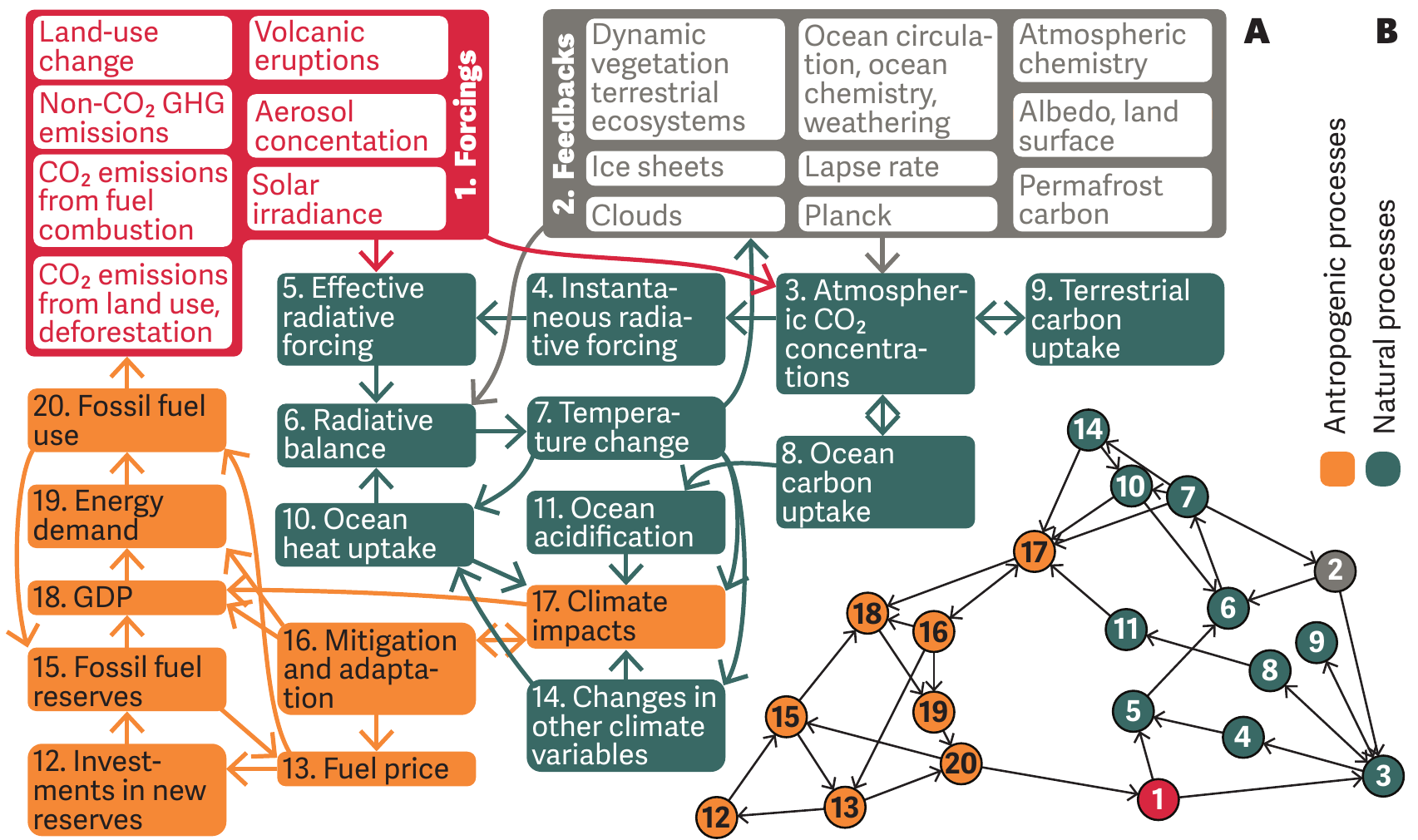}
  \caption{\textbf{The causal structure of a climate model.} The network in panel A, adapted from Ref.~\cite{knutti_feedback}, illustrates some components and their interdependencies in a typical large-scale climate simulation model. In panel B we show the same network as a directed graph where we keep the color coding of the initial illustration.}\label{fig:circulation_model}
\end{figure*}

\section{Introduction}

Stating the obvious, studying climate change is challenging because it is so complicated. Not only are climatological factors coupled into a complex causal web, but humanity's response to this challenge also depends on a multitude of entangled processes~\cite{odum_odum,donges_technosphere,schellnhuber_earth,claussen_intermediate}. Looking closer at any component of this network of causes and consequences, there are yet more networks: The photochemical reaction network of the atmosphere determines the concentration of greenhouse gasses~\cite{centler_photochemistries}. The in- and outflows to this system are affected by the bio-, geo- and technospheres' chemical networks~\cite{lal_carbon}. The human contribution of greenhouse gases depends on the opinion dynamics of climate action propagating over a web of people~\cite{dunlap_climate,farrell_counter}, and the global economy, which is, of course, a network~\cite{schweitzer_economic}. (See Fig.~\ref{fig:overview}.)

From the above observations, it is fair to call the science of climate change a network science. However, it is not the only such: From power grids to the neural system; from the narrative structure of novels to subway systems; from friendship contacts to money-flows in banks; all these systems can be understood as networks~\cite{newman_book}. The function of these systems is, to some extent, determined by the way they are connected~\cite{newman_book}. Thus, a first step to analyze them is representing them as graphs---mathematical objects consisting of nodes pairwise connected by edges. To extract meaningful information about a system from a graph is a rapidly developing, interdisciplinary field~\cite{newman_book,barabasi_book}. Once you can represent a complex system by a network, you can choose from an abundance of methods to analyze it. In this review, we will survey applications of network theory to the science of climate change and discuss the future of integrating network modeling into climate research.

On the one hand, as we will argue, network theoretical methods have entered the study of climate change late and are yet to be employed to their full potential. On the other hand, networks are close to the heart of climate science. Today's network science carries the legacy of 1970's systems theory~\cite{odum_odum,forrester1971world}, which is also an antecedent of much of today's climate modeling~\cite{edwards_history}. The similarities are easy to see, starting from the emphasis on an integrative, holistic view. The main difference is perhaps that network theory stresses \emph{how} things are connected. In Fig.~\ref{fig:circulation_model}A, we show a diagram of the components and interdependencies of a state-of-art climate model (adapted from Ref.~\cite{knutti_feedback}). In Fig.~\ref{fig:circulation_model}B, we plot the same diagram by a force-directed algorithm to highlight its network structure. This network will be a standing example of our paper.

In the remainder of this review, we start with an expos\'e of the tools and techniques of networks science and the questions they are designed to answer that pertain to climate change. Then we will review how these tools have been used in the science of climate change, organized by topics roughly sorted from causal to consequential (even if we recognize that is not strictly meaningful, cf.\ Fig.~\ref{fig:overview}). We aim to cover all issues directly related to the causes and consequences of climate change. The area just outside of our scope includes network science of sustainability~\cite{sayles_social}.

\section{Topics and tools of network science}

The mathematics of networks is a common language to describe the large-scale structure of vastly different systems throughout the natural, social, and formal sciences. With a network representation of a system of interest at hand, there is a multitude of analysis methods available~\cite{newman_book,barabasi_book}. This section will overview these methods, and explain how to adapt these methods to specific needs. It is organized around four typical questions that network science can help to answer. Code for the non-standard methods we use, and the data, can be found at \url{https://github.com/pholme/ncc}.

\subsection{Identifying important nodes and edges}

Once we have represented a system as a network, one approach is to zoom into individual nodes and edges and investigate their function. Often one would seek a ranking of the nodes or edges in order of importance. These would then be a priority list of what nodes to protect (in, e.g., an ecological network~\cite{allesina_pagerank}) or influence (in, e.g., a network of stakeholders in greenhouse-gas-emitting industries~\cite{bergsten_gaps,liu_systems,sayles_social}).

The meaning of importance ultimately depends on several factors: what the network represents, what dynamics are coupled by the network, how one would like to influence the system, and what kind of interventions are at hand. However, even though there are innumerable answers to these questions, the measures typically fall into a few different categories: centrality, vitality, and controllability measures.

\begin{figure*}
  \includegraphics[width=0.9\linewidth]{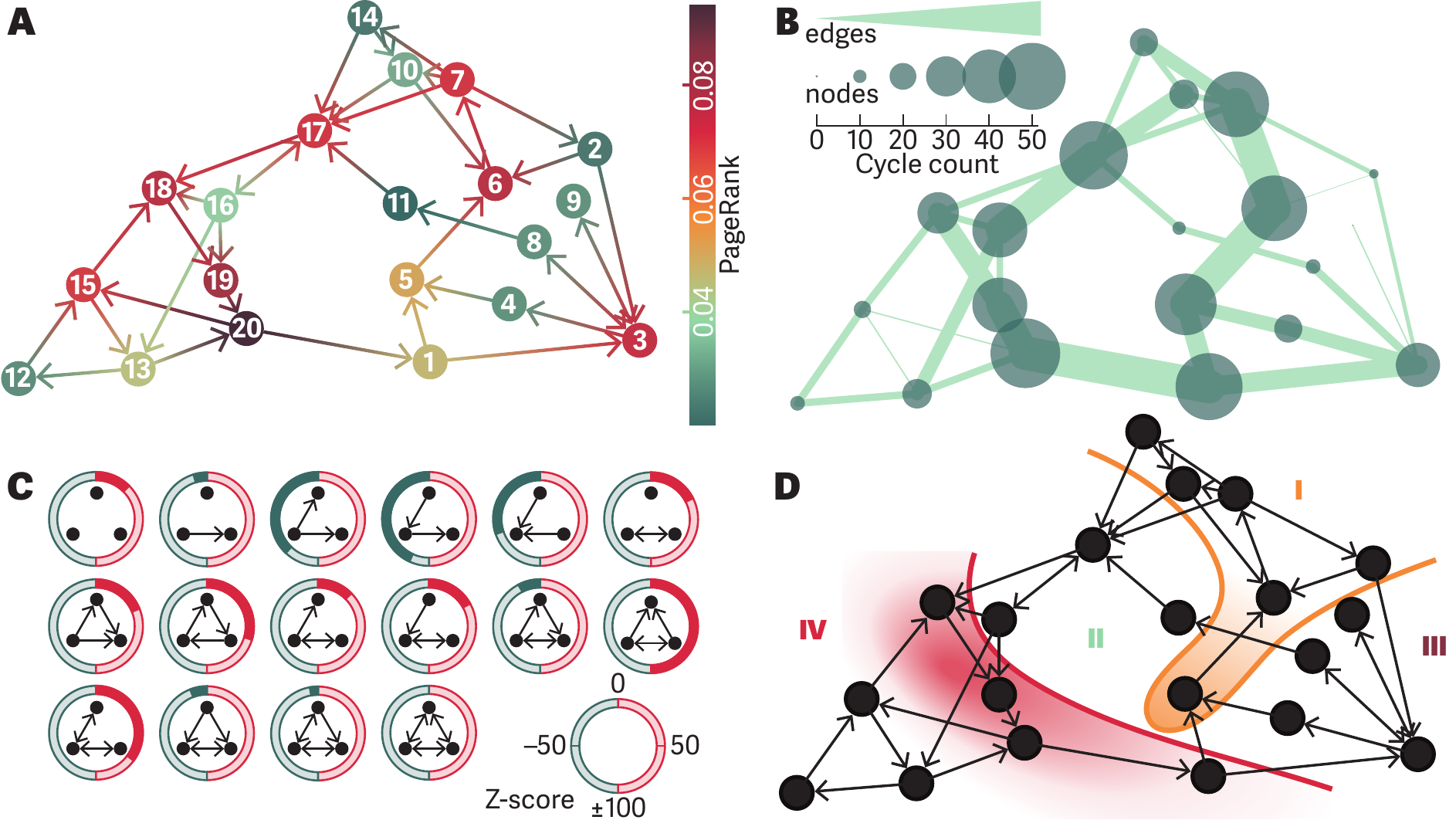}
  \caption{\textbf{Some applications of network-science methods to the network in Fig.~\ref{fig:circulation_model}.} Panel A shows the PageRank values of the nodes by color. Panel B shows counts of simple cycles by size for both nodes and edges. Panel C illustrates a motif analysis. We show all isomorphically distinct three-node subgraphs and their Z-score with respect to a degree-preserving null model. A large Z-score means that the subgraph is overrepresented in the original graph. Panel D uses the community detection algorithm InfoMap~\cite{infomap} to split the network into well-defined clusters.}\label{fig:netsci_showcase}
\end{figure*}

The simplest importance measure is the \textit{degree}---the number of neighbors of a node. As the name suggests, \textit{centrality measures} interpret a network as a geometric object and use distances and imaginary flows to rank the nodes. One prominent example is PageRank---the approach to ranking webpages that became the starting point for Google~\cite{barabasi_book}. Assume a walker jumps from node to node across the network. With a probability $\alpha$ (traditionally $\alpha$---the ``attenuation factor''---is set to 0.85), the walker follows a random link out from the node where it is. Otherwise, the walker goes to any node of the network at random. The PageRank of node $i$ is then proportional to how often the walker visits $i$. PageRank has been applied to directed networks far beyond web pages, such as identifying keystone species in foodwebs~\cite{allesina_pagerank}. Fig.~\ref{fig:netsci_showcase}A shows the PageRank values of the network in Fig.~\ref{fig:circulation_model}.

Another common approach to centralities is to look at the shortest paths between pairs of nodes in the network. The \textit{betweenness centrality} of node $i$ is essentially the count of shortest paths between all pairs of nodes that pass through $i$. Several works related to climate change have used this measure to rank nodes~\cite{rocha_cascading,farrell_counter,kronke_dynamics,donges_backbone,donges_complex}.

To illustrate the kind of reasoning that goes into the development of new centrality measures, let us modify betweenness to suit better the purpose of climate models such as the one shown in Fig.~\ref{fig:circulation_model}. Such models have always emphasized cycles~\cite{edwards_history}---closed paths, rather than shortest paths between distinct node pairs. So, rather than counting the latter, we can consider all the \textit{simple cycles} of the graph---cycles where a node appears at most once. Fig.~\ref{fig:netsci_showcase}B shows the counts of simplest cycles passing through the nodes and edges of our climate-model dependency graph.

Both PageRank and our cycle betweenness rank node 20 (fossil fuel use) highest, but some other nodes are ranked remarkably differently. Node 15 (fossil fuel reserves) has a high PageRank but low cycle betweenness, whereas the situation is reversed for node 16 (mitigation and adaptation). This shows that one needs to be careful to choose a centrality measure that is readily interpretable in one's investigation. Better still would be to validate centralities before picking one---i.e., to pick the measure with the highest predictive or explanatory power regarding some observed or simulated outcome.

\subsection{Mesoscopic structures (modules, motifs, clusters, or communities)}

If we zoom out from the node- and edge-centric perspective of importance measures, we come to mesoscopic structures. At this scale, network methods typically concern finding subgraphs that could represent functionally or conceptually clear subsystems.

Methods to discover network motifs are a common mesoscopic analysis tool (see Refs.~\cite{kronke_dynamics,lubell_network,jasny_echo} for applications to climate change). Motifs are statistically overrepresented subgraphs compared to the average of a null model~\cite{alon_motif}. Often one can interpret motifs analogously to components of electronic circuitry. For example, the three-node subgraph of nodes 1 and 2 both pointing to 3 and 1 also pointing to 2, could function as a logical AND-gate~\cite{milo_motif}. To keep the calculation fast and the result interpretable, motif analysis is often restricted to subgraphs of just three or four nodes.

A crucial step when identifying motifs is to choose a relevant null model. To make the motifs meaningful, the model should sample all graphs with the same basic constraints as the original graph with equal probability. The most common null model are graphs with the same number of nodes and edges and the same degree profile (i.e., the same number of nodes with in-degree $k_i$ and out-degree $k_o$) as in the original graph. Since our example graph represents a circulation model constructed to be strongly connected (there is a path from every node to every other node), we add the constraint of strong connectivity to our null model. We display the result for three-node subgraphs in Fig.~\ref{fig:netsci_showcase}C. The most overrepresented subgraph consists of two mutually dependent nodes that both influence a third node. Even though there is only one such subgraph---16 (mitigation and adaptation) and 17 (climate impacts) influence each other while both pointing at 18 (GDP)---such a configuration is so rare in the null model that the motif is highly unlikely to appear by chance. Straight feedforward paths like 3 (atmospheric CO$_2$ concentrations) pointing to 4 (instantaneous radiative forcing) pointing to 5 (effective radiative forcing) are so common in the null model that it comes out as an anti-motif (a subgraph suppressed by the network-forming forces). Presumably, this could be because the middle node could be merged with either of the end nodes when building a climate model.

Another mesoscopic methodology in network science is to detect \textit{communities}~\cite{prager_network,peter_bayes,silva_atmospheric}---loosely defined as subgraphs with an intricate inner circuitry and relatively few in- and output terminals. There is a vast number of methods for this purpose. The probably most common one, the \textit{Louvain method}~\cite{louvain_method}, heuristically optimizes the \textit{Newman-Girvan modularity}~\cite{newman_book}---a measure of partitions of graphs that increase if many links are within the partitions rather than between them. In Fig.~\ref{fig:netsci_showcase}D, we show the result of another approach, \textit{InfoMap}~\cite{infomap}, building on an information-theoretic idea that algorithms could use relevant communities to compress the description of random walks on the network. From the network structure alone, this method roughly recreates the anthropogenic component of the original graph as one community (IV), whereas it splits the rest of the network into three communities.

\subsection{Mechanisms driving the network evolution}

Real-world networks are rarely completely random. Instead, they have regularities, network structure, that can tell us something about the system's evolution.  We have already mentioned one such observation and explanation---that the suppression of feed-forward chains in our example network might come from typical considerations of model building.

A common approach is to look at quantities at a network-wide level---thus completing our zooming out from node-centric analysis via mesoscopic structures to global properties. The most common such inquiry is the study of the probability distribution of degree~\cite{kronke_dynamics,silva_atmospheric,sugiarto_regime,tsonis_architecture,tsonis_teleconnections,sole_reaction}. Such are often is heavy-tailed in empirical networks~\cite{barabasi_book}, and there are several mechanisms proposed generating such structures. Fat-tailed degree distributions are only one example of an interesting global network measures; see Refs.~\cite{newman_book,barabasi_book} for more examples.

Finally, we note there are methods of going straight to inferring the growth mechanisms of networks without trying to reproduce a known network structure~\cite{overgoor_choosing}. Such methods are yet unused in the climate change literature.

\subsection{How network structure affects the system dynamics}

Another common research question in network science is to relate the structure of a network to some dynamics occurring on the network (like opinion dynamics---to take an example from the climate change literature~\cite{jasny_echo}). Social networks typically have an overabundance of triangles~\cite{kossinets_empirical} which tend to slow down the spreading of opinions or information than spearing on a random network of a similar size. Assume A spreads climate-denial sentiments to B and C. Then, if B was connected to another node, D, the spreading would have been more efficient since D would not necessarily have the same opinion as B.

\subsection{Finding a method that suits your project}

How network methods enter a research project depends both on the available data and research questions. Often one is faced with a choice between using simpler or more complex network representations. Suppose one decides to include information about directionality, weights of nodes or edges, how these vary in time, etc. Then the predictive and explanatory power should, in principle, increase, but the number of available methods that can handle the complexity of the data would be much fewer. Often, however, it is reasonably straightforward to extend a concept to suit your needs---like our cycle betweenness (Fig.~\ref{fig:netsci_showcase}B) or the specialized null-model when calculating motifs (Fig.~\ref{fig:netsci_showcase}C).

Finally, network methods do not exclude other approaches)---regression analysis, differential-equation modeling, machine learning, etc. One would often use network methods to complement, or as a complement to, other methods.

\begin{figure*}
  \includegraphics[width=0.9\linewidth]{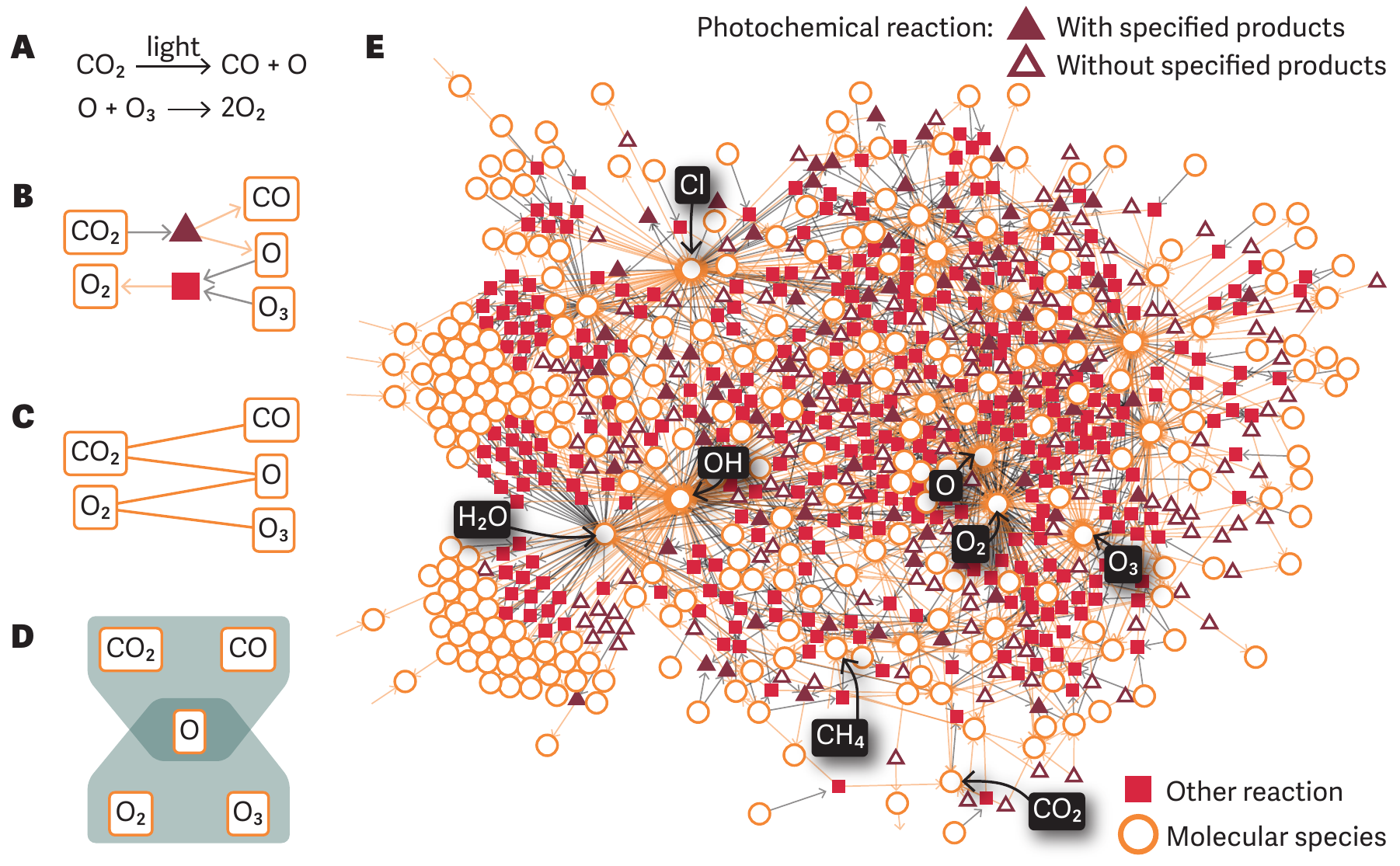}
  \caption{\textbf{Network structure of Earth's atmospheric reaction system.} Panel A shows a minimal reaction system comprising of two reactions. Panel B illustrates a representation with two types of reaction nodes (photochemical and non-photochemical). Panel C shows the most common projection to a simple graph. Panel D displays a higher-order network representation~\cite{battiston_beyond} where the molecular species involved in one reaction form a hyperedge---a generalization of edges to represent the interactions of arbitrary number of elements. Panel E shows the atmospheric reaction network of Earth (data from Ref.~\cite{sole_reaction}) in a representation similar to panel B.}\label{fig:atmoschem}
\end{figure*}

\begin{figure}
  \includegraphics[width=\linewidth]{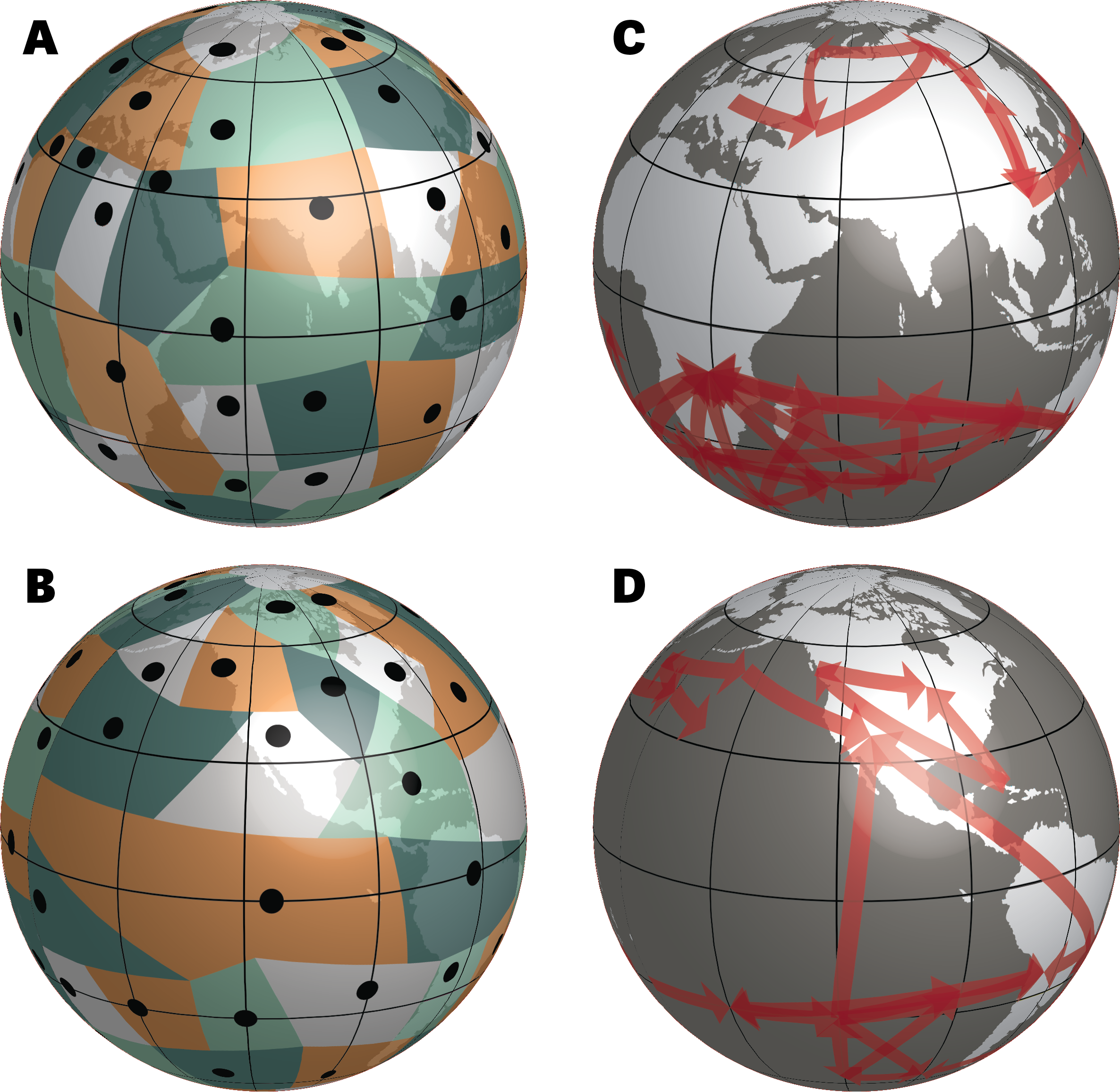}
  \caption{\textbf{Teleconnections in the Earth's climate.} The figure is adapted from Figs.~1 and 7 of Ref.~\cite{hlinka_directed}, and slightly simplified. Panels A and B show regions where the ground temperatures are highly correlated. The dots marking the most representative grid point. Panels C and D show Granger causality links between these regions. I.e., an arrow from region $i$ to $j$ means that the temperatures at $i$ are correlated to those at $j$ if they are shifted forward in time. We omit regions without arrows. }\label{fig:weather_dependencies}
\end{figure}

\section{Causes: Flows of matter, energy, and opinions}

A simple explanation of the causes of global warming does not need more than a sentence---gases injected by humans into the atmosphere shield the thermal radiation into space, raising the Earth's mean temperature. However, going just one level deeper, we are hit by a wall of complexity, as alluded to above. In this section, we will take a network view of these explanations.

\subsection{Global circulation models}

We start our discussion of concrete networks with large-scale climate models that inform the Intergovernmental Panel on Climate Change (IPCC) and decision-makers worldwide. These are primarily simulation models made with prediction and scenario testing in mind. They are built around a dynamic simulation of fluid motion across a physical model of the Earth~\cite{edwards_history}. This fluid simulation is coupled with other physical feedback mechanisms and sometimes social components (Fig.~\ref{fig:circulation_model}). These couplings are causal---changing a variable in one end of a link can change the state of the node at the other end. The causation could happen through meteorological flows (of matter, energy, or some other thermodynamic entity), economic couplings, or political influence.

Some design principle of general circulation models are:
\begin{itemize}
    \item Increasing accuracy by adding missing components and increasing spatial and temporal resolution.
    \item Increasing computational performance and eliminating computational bottlenecks~\cite{washington_petaflop}.
    \item Increasing extensibility and facilitating validation by standardizing the components of the models.
\end{itemize}
Simplifying these models has never been a main objective, meaning that they have, after over 60 years in the making, reached a degree of complexity beyond what a single research group or even a single organization could handle~\cite{kronke_dynamics}. This also means that they represent, relatively completely, our mechanistic understanding of climate processes. Furthermore, since these models are meticulously validated~\cite{flato_evaluation}, they constitute proofs of concept that few other theories, in all of science, can match~\cite{holme_mechanistic}.

Critics may say that general circulation models are so complicated that they no longer represent a scientific explanation~\cite{claussen_intermediate}. Understanding the models (and thus climate) calls for a network analysis, which closes the loop to the systems diagrams of yesteryear~\cite{odum_odum}.

\subsection{Chemical reaction systems}

The primary cycles of atmospheric greenhouse gasses are part of much larger chemical reaction systems~\cite{centler_photochemistries}. To understand the organization of these larger systems, one can use network approaches~\cite{silva_atmospheric,sole_reaction}.

Fig.~\ref{fig:atmoschem} shows a chemical reaction network of the Earth's atmosphere. Note that this network is integrated with the chemical processes of the oceans and land. When it comes to CO$_2$ capture, these other global chemistries could aid the biosphere~\cite{lal_carbon}. Given the versatility of whole-system network modeling in the `omics disciplines and biochemical modeling~\cite{barabasi_book,newman_book,alon_motif}, it would be interesting to construct a database with all reactions on Earth, outside of the biosphere.

There are several established ways of analyzing chemical reaction systems. These can incorporate more chemical constraints than a pure network-science approach. For instance, \textit{flux-mode analysis}~\cite{orth_flux} is a technique assuming both an underlying reaction network and additional restrictions from mass conservation, but not reaction kinetics.

It is worth mentioning that chemical reactions lend themselves well to \textit{higher-order network representations}~\cite{battiston_beyond}---one of the hottest topics at the moment in network science. In such abstractions, more than two nodes can interact at one time. For chemical reactions, we can, e.g., use this to encode the restriction that molecules of two different species need to meet at a point in space and time for a reaction to occur. These directions are, at the time of writing, largely unexplored.

\subsection{Climatic teleconnections}

The presence of long-distance correlations in weather patterns has been known since the 19th century~\cite{teleconnections}---the emblematic example being the El Ni\~no Southern Oscillation (ENSO) that connects weather in the equatorial Pacific Ocean. Rather than studying the full correlation patterns, one can reduce them to a network~\cite{steinhaeuser_complex}. A cell in such a network typically represents a region of relatively cohesive weather patterns. Different cells are linked if the condition in one cell predicts the weather (technically authors use pressure~\cite{tsonis_architecture,tsonis_teleconnections}, temperature~\cite{fan_localized,gozol_elnino,yamasaki_elnino}, rainfall~\cite{boers_teleconnections}, or sea currents~\cite{dijkstra_networks}) in the other~\cite{tsonis_architecture}. Since we can expect the weather dependency patterns to change because of global warming, understanding the dynamics of climate networks could inform climate models and improve our forecasting of climate change~\cite{fan_weakening}.

Once we have established the links of such a climate network, we can describe the dynamics of a climatic episode~\cite{gozol_blinking}. For example, Gozolchiani \etal\ characterized the El Ni\~no phase of ENSO by the El Ni\~no basin~\cite{gozol_elnino}, at the onset, losing its influence on the rest of the networks. After some months, the situation reverses, so a large part of the weather network becomes influenced by the El Ni\~no basin.

In Fig.~\ref{fig:weather_dependencies}, we show an example of a network of Granger causality of weather patterns from Hlinka \etal~\cite{hlinka_directed}. Several papers on this topic focused on how to derive meaningful networks of weather dependencies~\cite{fountalis_spatio,imme_graphical,imme_new,donges_backbone,donges_complex,hlinka_directed,runge_causes,runge_climate,steinhaeuser_complex}, others showed that climate networks reflect well-known features of synoptic scale meteorology~\cite{wang_rossby,yamasaki_elnino,sonone_elnino}, or that climate network can help in predicting extreme events~\cite{boers_extreme,boers_teleconnections,malik_monsoon}. Fan \etal~\cite{fan_localized}, took the weather dependency network modeling one step further and considered a network where nodes are El Ni\~no events, linked by their similarity. Steinhaeuser \etal~\cite{steinhaeuser_multi} also presented a relatively unusual climate network study in that it builds on multilayer networks---networks where there are different classes of nodes and edges~\cite{newman_book}.

\subsection{Managing common-pool resources}

We can understand many kinds of climate decisions through game theory. Even though humans are not necessarily well-informed and relentlessly selfish---as typically assumed in economic modeling---game theory can clarify what is at stake in a particular decision.

Many goods related to climate change are \textit{common-pool resources}. This means they are non-excludable (in principle, everyone has access to it) and rivalrous (the used part of the resource becomes unusable or degraded for the future). Most game-theoretical studies do not rely on networks, but some do. When agents interact through a social network, the network structure can substantially affect the dynamics. For example, Chung \etal~\cite{chung_coupled} showed that once the social network compels people to follow norms (like conserving common-pool resources), more social interaction promotes cooperation. This finding was explored in a dynamic setting by Sugiarto \etal~\cite{sugiarto_regime} who found that such norm-enforcing social interactions could create a hysteresis effect and thus the possibility of sudden behavioral (and related ecological) changes. On a positive note, social networks could be susceptible to governmental interventions~\cite{sugiarto_2017}, even though they fail to manage a common pool resource when operating in isolation. Furthermore, systems of reinforcement learning agents could also sustain limited shared resources~\cite{ml_cpr}.

\section{Consequences: Impact on nature and society}

Now we turn to studies focusing primarily on the consequences of climate change. Such consequences are, of course, coupled back to the causative factors, and all cited studies duly regard themselves as contributing to our understanding of a minor part of this grand cycle.

\subsection{Impact on ecological network structure}

Traditionally, ecologists have used networks to study species interactions such as predation or mutualism (e.g.\ pollination). Network representations of ecological systems have helped assess the potential impacts of climate change on community assembly and its stability~\cite{morris_forest}, both in modeled~\cite{saavedra_tolerance} and empirical mutualistic networks~\cite{bascompte_mutualism}, as well as foodwebs~\cite{gilarranz_foodwebs}. We can expect climate change to affect the niche and geographical range of species distributions and, with them, the strength of species relationships. Network science has been useful to predict the probability of secondary extinctions given climate disturbance~\cite{bascompte_mutualism} or the effect of warming in fitness depression in host-associated microbiomes~\cite{greenspan_warming}. Climate change can cause a shift from bottom-up to top-down regulation of herbivorous communities in Arctic foodwebs~\cite{legagneux_arctic}, which implies tracking the change over time of link strength in ecological networks. Furthermore, Gilman \etal~\cite{gilman_community} used a very detailed model of ecosystem response to climate change and found that ecosystems with more links were more robust to global warming. In a similar vein, Schleuning \etal~\cite{schleuning_change} studied pollination and seed-dispersal networks with over 700 species. In a hybrid approach---with climatic niche-breadth estimates coupled to simulations---they concluded that ecological networks are more sensitive to plant than to animal extinction under climate change.

\subsection{Tipping points in ecological networks}

Networks have also been used to analyze tipping points in ecological systems, their interactions with climate, and potential cascading effects.

Ecosystems at large scales can undergo pervasive and abrupt changes if they transgress tipping points. We can expect climate change to increase the rate of such critical transitions~\cite{bolt_redding}. For example, the shift from rain forest to savannas has been estimated from remote sensing data to have a tipping point at around 1500 mm of precipitation per year~\cite{verbeset_slowness} or reduction of more than 40\% of area. Climate change projections for marine ecosystems predict unprecedented abrupt changes in global oceans~\cite{beaugrand_prediction}. The climate system has at least 15 of these tipping elements identified~\cite{lenton_risk, lenton_tipping}, and the list readily grows to over 30 types of critical transitions if we include local and regional ecosystems~\cite{rocha_cascading}. While climate-related processes (e.g., temperature increase, changes in precipitation, or drought) are one of the main causes of irreversible change in ecosystems, it is seldom the only one. A network study that exploited the bipartite structure of drivers and regime shifts showed that climate-related variables often interact with food production (e.g., agriculture, fishing), changes in nutrients flow, and urbanization~\cite{rocha_drivers}.

Some systems with tipping points have connectivity features where networks can enrich the analysis. For example, shifts from rainforests to savannas are mediated by moisture recycling feedbacks, i.e., rain transport dynamics from one place to another. Networks have been used to make such connections explicit in space and time. For example to assess how regime shifts in a region of the Amazon can affect the likelihood of regime shifts in another area~\cite{kronke_dynamics}.

Tipping points have also been hypothesized to be interconnected across different climate~\cite{lenton_risk} and ecological systems~\cite{hughes_multiscale, steffen_hothouse}. The network literature already offers some tools that might be adequate to address potentially cascading tipping effects. For example, an increase in connectivity and a decrease in modularity can amplify tipping cascades~\cite{brummitt_cascades, brummitt_suppressing}. Ref.~\cite{kronke_dynamics} showed that some network motifs can destabilize interconnected tipping elements, and high clustering (many triangles) exacerbates the systemic vulnerability to regime shifts. Feedforward loops can decrease the coupling strength necessary to initiate cascades, a finding applied in models of the Amazon rainforest~\cite{wunderling_motifs} and some climate tipping points~\cite{wunderling_interacting}. Yet, these models assume a small number of systems and a known coupling mechanism. The analysis of high-order motifs on causal networks, such as feedback cycles, has enabled the exploration of alternative mechanisms that couple tipping elements~\cite{rocha_cascading}.

We still lack empirical evidence of cascading tipping points~\cite{scheffer_seeing}, which is thus one of the most pertinent open questions. Combining network-based causal inference methods with climate simulations or Earth system observations are likely fruitful avenues for future research.

\subsection{Impact on socio-ecological systems}

Predicting anthropogenic responses is perhaps the most challenging task in climate science. That much said, it is maybe not hopeless---sometimes humans behave very predictably. Mitigating climate change is maybe not hopeless either---sometimes humans collectively solve emergencies with exceptional efficiency.

Models of human response to climate change are very diverse, but many focus on feedback effects from the climate to human behavior. For example, Peter \etal~\cite{peter_bayes} studied how a warmer climate would affect biofuel production. They construct a complex Bayesian model of South African agriculture, including around a hundred variables (like ``soybean water use'') and their dependencies. Yletyinen \etal~\cite{yletyinen_multiple} used a simulation of ecological decision-makers to argue that oversimplifying social interactions be gravely misleading about environmental outcomes. A yet more complex simulation platform, including networked human decision-makers and ecosystems alike, was presented by Donges \etal~\cite{donges_esd}. We note that the authors call the model simple, and given all components they seek to include, it is a fair description. Still, compared to, e.g., the networked game-theoretical models discussed above, it is very intricate.

Several studies concern the impact of global warming on the management of limited resources. Prager and Pfeifer~\cite{prager_network} took a network approach to understand rainwater management among Ethiopian smallholders and its resilience to climate change. The authors constructed a multilayer network representation with separate nodes for people and plots. Links in the plot-plot networks represented water flow from one field to another. The social network of smallholders reflected the fields in that downstream farmers would depend on their upstream neighbors. Mina \etal\ investigated the management of forest by a combined spatial simulation and network analysis~\cite{mina2021network}. Many studies used similar social-network approaches to understand the resilience of governance of shared resources, but less directly related to climate change. For example, Jansen \etal~\cite{janssen_toward} used a network approach to analyzing irrigation in Darwin, Australia, and Lubell, Robins, and Wang~\cite{lubell_network} employed exponential-random graph models~\cite{ergm,newman_book} to understand water management in the Bay Area, California.

A vastly different type of social impact are population displacements in response to global warming~\cite{migration}. Catteano~\cite{cattaneo_migrant} pointed at the role of exiting social networks in the demographic changes from people seeking environmental refuge likely ahead of us---a paper without explicit network analysis but hinting at the need for such. Researchers have also studied the climate-change impacts of networked infrastructures like roads~\cite{galbraith_scottish,chinowsky2013assessment} and power grids~\cite{mccoll_power}. However, so far, without explicitly quantifying the implications in terms of network structure.

\subsection{Opinion dynamics}

Climate action is a consequence of social information spreading, both in the networks of stakeholders and politicians and among the general public. In other contexts, social influence over networks of friendships is a well-studied topic~\cite{lehmann_complex}. Typically, articles either present analyses of social media data~\cite{cann} or simulation studies~\cite{sugiarto_2017,sugiarto_regime}.

In network studies of social-media posts about climate change, the nodes are individuals, and links represent information sharing from one person to another. We will mention four examples all drawing data from Twitter. The boundaries of the data set then define the scope of the study. For example, Cann \etal\ analyzed a Twitter network to understand public sentiments around the announcement of the US withdrawal from the Paris Agreement. Using community detection~\cite{louvain_method}, they discovered a binary split (although far from complete) into supporters and antagonists of the US action. Vu \etal\ \cite{vu} also used Twitter data, but (rather than gleaning tweets about a specific topic within a particular time window) they gathered the follower network within a set of NGOs in the climate change sector. Their analysis focused on employing centrality measures to identify key actors. Tyagi and coworkers~\cite{tyagi2020affective}, used a richer network representation---a so called ``signed network'' where positive links represent agreement and negative represent antagonism. They could characterize polarization by analyzing the boundary represented by negative links. In our final example, Goritz \etal\ argue that one can use network analysis of discussions on social media in teaching topics on climate change~\cite{su11195499}.

\subsection{Policy networks}

 The study of policy networks is a well-established field within political science. A policy network consists of actors---organizations or individuals---linked by common interests or beliefs concerning policymaking~\cite{policy_network_analysis,kim}. Whereas the actors are typically officially recognized entities, links could be anything from informal to institutionalized.

How the scientific information about climate reaches decision-makers can be staggeringly complex. Broadbent and Vaughter investigated how information the IPCC enters the information flow of policy networks in New Zealand and Japan~\cite{broadbent_sna}. In a similar study, but focusing on information spreading by climate denialists, Farrell~\cite{farrell_counter} used official statements, press releases, and news articles to study the network of information spreading intended to hinder climate action. Furthermore, by employing topical modeling techniques from natural language processing and network centrality measures, Farrell concluded that top corporate actors are the drivers of this movement.  In addition, he noted a trend of actors deliberately obscuring this network. A few papers studied the network of information sharing among policy actors~\cite{jasny_echo,williams_echo}. Using motif analysis (presented above), they found evidence of echo chambers that effectively align the information that reaches the actors. If the communication happens between agents of different viewpoints, to could, however, mitigate some environmental crises~\cite{barnes_outcomes}.

Researchers have also used questionnaires to map the connections between policymakers. Howe \etal focused on the Canadian climate policy network~\cite{HOWE2021}. Using an exponential random graph framework~\cite{newman_book}, they studied how the position of actors in the network predicted their stance on specific issues like oilsand exploitation. Lee and van de Meene used the same approach---analyzing questionnaire data by exponential random graphs---to elucidate how cities learn climate action from each other~\cite{lee_meene}. Yl\"a-Antilla \etal\ argue for an international perspective for interorganizational policy network research on climate change comparing Canadian and Finnish data~\cite{YLAANTTILA2018258}. Papin~\cite{papin} and Bansard \etal~\cite{bansard} take an international perspective on networks of cities and their policies with respect to climate change.

Policy networks are coupled to financial networks. Stolbova, Monasterolo, and Battiston used economic theory and data to analyze how shocks of new climate policies propagate through financial networks~\cite{STOLBOVA2018239}. Brulle studied another type of financial network---the network of money flow among organizations of the climate change counter-movement in the USA~\cite{brulle_institutionalizing}---concluding that the most central companies are the drivers of this system. I.e., the same conclusion reached by other methods by Farrell~\cite{farrell_counter}.

\section{Discussion}

\subsection{Summary}

For over a decade, climate scientists have employed network methods---from governance studies via ecological tipping points to the web of climatic teleconnections. To network scientists from other backgrounds, this seems obvious. The second of the \textit{Essential principles of climate literacy} by the U.S. Global Change Research Program states that ``climate is regulated by complex interactions among components of the Earth system.''~\cite{climate_literacy} This almost directly portraying the Earth system as a complex network, so naturally we need complex-network methods to understand it.

To a mainstream climate-change or Earth-system scientist~\cite{schellnhuber_earth}, network science will feel like the familiar stranger you see on your commute every morning. All these fields share a great deal of intellectual ancestry. All recognize the failure of reductionist science to account for emergent phenomena like global warming and the human response to it. They also differ from machine-learning applications in that they rely on causal arguments rather than learning from examples. The latter is, of course, challenging when it comes to dealing with the current (and only known) anthropogenic global warming. On the other hand, we need data science too, as it is arguably a powerful tool for several questions related to climate change~\cite{sugihara_detecting}.

Climate science has had a strong causal grounding due to its mechanical modeling tradition and comprehensive estimates of uncertainty~\cite{flato_evaluation}. Climate data, whether Earth system observations or scenario modeling output, usually comes as high dimensional arrays that lend themselves for further exploration of causal interdependencies of the climate system~\cite{runge_causes}, and its interaction with the biosphere~\cite{Krich_biosphere}, technosphere~\cite{donges_technosphere}, or social systems~\cite{Otto_social}. Modern methods for causal inference offer new opportunities to explore causal connections by reducing the high dimensional space to a directed network~\cite{sugihara_detecting,runge_causes, Krich_biosphere, runge_climate}. Caution is needed, however, to ensure that observations are long enough to capture the time scales of the dynamics or that the resolution captures the fast and slow processes of the interactions of interest. The interplay between climate and other spheres are fruitful areas of research, where networks as boundary objects are likely to play an important role. Networks can, thus, enter causal modeling in many ways: the input data could be a network (e.g., networks of information flow), the output could be a network of causal connections of the same types of units (e.g., climatic teleconnections), or a causal network connecting model components to a greater whole (cf.\ Fig.~\ref{fig:circulation_model}).

\subsection{Limitations of network approaches}

Network-based methods are tools in the toolbox, not an entire toolbox. The problems that one could solve with a pure network approach are limited to those where the only available data is a network of binary, possibly weighted, interactions (e.g., policy networks). When other information is available, a network approach would typically also involve different types of modeling. For example, a differential equation system could be described by the network (leading to systems dynamic type models), or the network could result from spatial data (like meteorological teleconnections). Indeed there is a full spectrum, from pure network problems to those where network methods only add a marginal value.

\subsection{Conclusions and future directions}

Applications of network science to climate change are still far from mature, and in all the topics mentioned in this review, there is more work to do. Recognizing the social factors behind many of the processes in play~\cite{blok}, we believe applications of social network analysis are particularly promising. One understudied issue is the impact of climate change on vulnerable populations~\cite{jones_ready}. Another interesting focus area is migration triggered by environmental deterioration~\cite{migration}. Social networks are thought to be a critical determinant of the migration flow, where international migration, in particular, is guided by existing social ties. Networked infrastructures could also be sensitive to climate change~\cite{kourtis,galbraith_scottish} and relatively rarely studied by the methods of network science.
Finally, we believe climate science itself and our understanding of climate changes in all its complexity also would benefit from network studies (cf.\ Fig.~\ref{fig:circulation_model}).

To summarize, network science has found its way into the toolbox of an increasing number of climate scientists. We envision that network approaches eventually become a network in itself, methodologically connecting different areas of the science of climate change.

\acknowledgments{The authors thank Onerva Korhonen and Milja Heikkinen for comments on the manuscript. P.H. was supported by JSPS KAKENHI Grant Number JP 21H04595. J.C.R. was supported by Formas grant no.\ 942-2015-731.}

\bibliography{earth}
\bibliographystyle{abbrv}

\end{document}